\documentstyle[12pt,manuscript,aps,epsf,colordvi]{revtex}

\input epsf
\newcommand{\frat}[2]{\frac{\textstyle #1}{\textstyle #2}}

\newcommand{\vf}[1]{\mbox{\boldmath $#1$}}

\begin{document}

%----------------------------

\begin{center}
{\Large \bf  Field strength correlators in the instanton gas:\\
The importance of the two--instanton contribution} 
\\
%--------------------------------------------------------
\vspace{0.4in}
{\bf E.-M. Ilgenfritz$^{1}$, B. V. Martemyanov$^2$,
 M. M\"uller-Preussker$^3$}\\ \vspace{0.4in}

$^1$Research Center for Nuclear Physics, Osaka University, Japan \\
$^2$Institute of Theoretical and Experimental Physics, Moscow, Russia\\
$^3$Institut f\"ur Physik, Humboldt-Universit\"at zu Berlin, Germany

\end{center}

\begin{abstract}
%===============
The field strength correlators at zero temperature are semi-classically 
evaluated fitting the random instanton liquid model to lattice data for 
quenched $SU(3)$ lattice gauge theory.
We restrict ourselves to the lowest order in an instanton density expansion 
necessary to explain the difference between ${\cal D}_{\bot}$ and ${\cal D}_{||}$.
In the instanton--instanton and instanton--antiinstanton contributions the 
Schwinger line factors -- neglected in a previous analysis -- are numerically 
taken into account in a weighted Monte Carlo evaluation.  This leads to different 
estimates for instanton size and density. A reasonable description of the 
correlators within the intermediate range from $0.4\,\rm{fm}$ to $1\,\rm{fm}$ is 
obtained.
\end{abstract}

\section{Introduction}
%======================
Based on the concept of gluon field-strength correlators, a model independent, 
albeit systematic description of non-perturbative effects in QCD 
\cite{simdosch,sim,simshev} has been worked out. Few correlation functions play an	 
important role in the stochastic confinement model and give a detailed description 
of the level splitting of heavy and light $\overline{Q}Q$ bound states \cite{quarkonia}.
Relying on the analytical continuation from Euclidean to Minkowski space, they 
provide a bridge between low and high energy QCD phenomenology: they are the basic 
ingredients for a description of high-energy hadron and quark-(anti)quark scattering
within the stochastic vacuum approach \cite{nachtmann}.

By now, numerical results are available from lattice simulations concerning the 
two-point field strength correlators for pure gauge theory with the gauge groups 
$SU(2)$ \cite{campos,dig1,ddeb} and $SU(3)$ \cite{dig2,dig3,dig4,bali} over physical 
distances ranging up to $1\,\rm{fm}$. The correlators have also been obtained near 
the deconfinement transition in pure $SU(3)$ gauge theory \cite{dig3}. 
Somewhat later, this study has been extended to full QCD with four flavours of 
dynamical staggered quarks \cite{dig4}.

In most of the recent lattice computations of the gluonic field strength correlators 
the signal has been extracted using a cooling procedure which serves the purpose to 
erase short--range fluctuations on the level of a few lattice spacings, thereby 
redefining (renormalizing) the field strength operator on the lattice. Apart from this 
renormalization, the correlator is expected to be not affected by this procedure as 
long as the diffusive cooling radius $r \propto \sqrt{n}$ remains small compared 
to the distances $d$ of interest (in units of lattice spacing) for an appropriate 
number $n$ of cooling iterations. Presently, this prevents the knowledge of the field 
strength correlator at very short distances below $0.1\,\rm{fm}$. Going to these 
distances requires to do such simulations on a few times finer lattices. 
In the finite temperature case and for QCD with dynamical fermions the correlators 
are presently known only for distances larger than $0.4\,\rm{fm}$.

With cooling like this, eventual semiclassical structures underlying the 
vacuum of Yang--Mills theory or full QCD can hardly be revealed from Monte Carlo lattice 
configurations. Nevertheless, one might try -- on the basis of the lattice data for the 
field strength correlators -- to justify and to constrain semiclassical models of vacuum 
structure. By definition, correlators formed by the field strengths of particular 
semiclassical configurations do not cover the perturbative part of what is measured on 
the lattice. In order to compare, one has to rely completely on the way how the lattice 
data are split into a (singular at short distances) lowest order perturbative part and 
the non--perturbative signal. This is the way how the renormalon ambiguity is presently 
dealt with (for a discussion see the recent talk by Di Giacomo \cite{DG_talk}). The 
non--perturbative signal is usually modelled by exponential contributions to the two 
basic structure functions ${\cal D}$ and ${\cal D}_1$ (see below). These can be replaced 
by other functional forms \cite{antonov}, {\it e.g.} suggested within high energy 
scattering phenomenology \cite{meggiolaro}, or provided by some specific vacuum model as 
in the present paper.

Instantons -- localized finite action solutions of the Euclidean Yang Mills field equations 
-- are well-known examples of semiclassical configurations \cite{thooft}, which have 
been put into hadronic phenomenology in a dilute gas model in Ref. \cite{cal}.  After 
the IR divergence had been cured in a rather {\it ad hoc} way taking interactions into 
account \cite{ilgmmp}, the instanton liquid model has been developed by Shuryak \cite{shur} 
and Diakonov \cite{diak}. For reviews we refer to Refs. \cite{diaktalk,shurschaf}.
Gradually, since instantons are now successfully identified in lattice ensembles (see the 
recent rapporteur talks \cite{negele,teper}), the instanton vacuum accumulates direct 
support from lattice QCD. However, there are still systematic uncertaincies concerning 
the parameters of the instanton liquid emerging from lattice analyses. 

Without doubt, instantons play an important role in explaining chiral symmetry breaking 
and other empirical facts of hadron structure. However, their contribution to the 
{\it confinement property} of non--Abelian gauge theory (as far as this is established on 
the lattice) turned out quantitatively insufficient \cite{fukushima,chen}, at least at the 
present level of analytically dealing with the instanton liquid model or implementing it in 
numerical simulation. One step beyond the current wisdom, {\it i.e.} beyond the neglect 
of instanton color correlations, has been attempted in Ref. \cite{ilgthur}.  The method 
proposed there uses the field strength correlator in a {\it dedicated way} between clusters 
of topological charge. However, based on the RG smoothing method, the instanton 
interpretation of the emerging clusters in \cite{ilgthur} has remained uncertain. Therefore, 
the same type of measurement should be applied in the context of any cooling study devoted 
to instantons. 

Relating instantons to the confinement issue, alternatively to the $\bar{Q}Q$ force,  
the property of monopole percolation in a multiinstanton system \cite{fukushima,monopolperc} 
or the effect of instantons on the vacuum structure as summarized in the field strength 
correlators  can be investigated. Earlier studies of field strength correlators based on the 
instanton liquid model can be found in Refs. \cite{ba,do1,IMMMPS}.  Here, there are two lines 
of thought. One \cite{do1,do2} relies on the one--instanton approximation and concentrates 
on the modification of the single instanton approximation (and the instanton solution
by itself) due to the interaction with the vacuum medium surrounding it.
{\it Our approach} tries to separate the contributions of one (anti)instanton from 
contributions when the field strength comes from the non--linear superposition of two 
(anti)instantons. Correspondingly, different results can be found in Refs. \cite{ba,do1,IMMMPS}. 

As far as the one-instanton contribution is concerned, we have nothing to add to our 
previous paper \cite{IMMMPS}. Concerning the two-instanton contributions, we had followed the 
current wisdom and neglected eventual color correlations in the instanton liquid. This is not 
a bad approximation for zero temperature quenched QCD, and we keep this approximation also now.
However, our results in the previous paper were obtained omitting the Schwinger-line phase 
factors in dealing with the two-instanton contributions. Some (rough) arguments led us to 
expect a negligible systematic error. This was correctly criticized in \cite{do2}, mainly 
because of some unavoidable consequences. The investigation to be presented here was 
triggered by this discussion and was aimed to improve our previous results avoiding this 
simplifying assumption.  This has led us into a rather involved weighted Monte Carlo 
treatment of the collective coordinate integration, taking the Schwinger line factors 
completely into account in the numerically evaluated integrand. It turns out that this 
treatment of the second order contributions partly changes our previous conclusions.

After recalling, for the sake of completeness, the notation for field strength correlators 
in Section II we concentrate on the discussion of the two--instanton contribution in Section 
III. For the first order results we can refer to Ref. \cite{IMMMPS} where they were obtained 
in an analytical way. Conclusions will be drawn in Section IV.

\section{The Field Strength Correlators in the Semiclassical Approximation}
%==========================================================================

The gauge invariant two-point correlators of the non-Abelian field strength are defined as
\begin{equation}
\label{eq1}
{\cal D}_{\mu\rho,\nu\sigma}(x_1-x_2)=
\langle 0|{\mathrm Tr}\{G_{\mu\rho}(x_1) S(x_1,x_2)
G_{\nu\sigma}(x_2) S^{\dagger}(x_1,x_2)\} |0 \rangle \,,
\end{equation}
where $G_{\mu\rho}= T^a G^a_{\mu\rho}$ is the field strength
tensor and $S(x_1,x_2)$ is the Schwinger-line phase
factor, {\it i.e.} the parallel transporter necessary to
join the field-strength operators at the points  $~x_1,~x_2~$
in order to respect gauge invariance.
$T^a$ denotes the generators of the gauge group $SU(N_c)$.
The most general form of the correlator compatible with Euclidean $O(4)$ rotational
invariance, adequate for zero temperature,  is in the notation of Ref. \cite{simdosch},
\begin{eqnarray}
\label{eq2}
{\cal D}_{\mu\rho,\nu\sigma}(x) & = &
\left(\delta_{\mu\nu}\delta_{\rho\sigma}
     -\delta_{\mu\sigma}\delta_{\rho\nu}\right)
\left({\cal D}(x^2) +{\cal D}_1(x^2)\right)+
\nonumber\\ [-.2cm]
&&\\[-.25cm]
&&+\left(x_\mu x_\nu \delta_{\rho\sigma}-x_\mu x_\sigma\delta_{\rho\nu}
+ x_\rho x_\sigma \delta_{\mu\nu}-x_\rho x_\nu\delta_{\mu\sigma}\right)
\frat{\partial {\cal D}_1(x^2)}{\partial x^2}\,,  \nonumber
\end{eqnarray}
with $x=x_1-x_2$ and ${\cal D}(x^2)~$ and ${\cal D}_1(x^2)~$
representing invariant vacuum structure functions.

\noindent
Instead of considering the invariant functions ${\cal D}(x^2)~$ and ${\cal D}_1(x^2)~$
entering Eq. (\ref{eq2}) we shall study longitudinal and transverse
correlators
\begin{eqnarray}
\label{dpar}
{\cal D}_{||}(x^2) & = & {\cal D}(x^2) + {\cal D}_1(x^2) + 
		 x^2\frat{\partial {\cal D}_1(x^2)}{\partial x^2} \,,
                              \nonumber\\
~~~\\
{\cal D}_{\bot}(x^2) & = & {\cal D}(x^2) + {\cal D}_1(x^2)\,.\nonumber
\end{eqnarray}

\noindent
In particular, if $x = (0,0,0,x)$ is Euclidean timelike, we have
\begin{eqnarray}
\label{dpar1}
{\cal D}_{||}(x^2) = \frat13 \sum_i {\cal D}_{4i,4i}(x^2)\,,\nonumber\\
~~~\\
{\cal D}_{\bot}(x^2) = \frat13 \sum_{i<j} D_{ij,ij}(x^2)\,.\nonumber
\end{eqnarray}

\noindent
It is easy to demonstrate that ${\cal D}_1~$ does not contribute to the 
area law of Wilson loops \cite{sim,simshev}.
In the perturbative regime both invariant functions
${\cal D}~$ and ${\cal D}_1~$ behave like $~1/x^4~$. Only  ${\cal D}_1~$
receives a contribution from one-gluon exchange. On very general grounds, the 
perturbative part of ${\cal D}~$ (which appears at one loop and higher orders)
was recently shown to be cancelled in the expression for the string tension 
by higher correlator contributions \cite{sim2}. Here, we shall not discuss 
the perturbative contributions in more detail. Instead we will concentrate 
on the contribution from instantons as exclusive semiclassical configurations 
representing the non-perturbative part of the correlators.

For comparison we refer to the lattice data published by the Pisa group 
for the two-point field-strength correlators in pure $SU(3)$ gauge theory at $~T=0~$ 
\cite{dig2}. Since these data include measurements for several bare lattice couplings 
$\beta$, the cooling method seems to be compatible with the correct scaling behavior 
at high $\beta$ of the structure functions.
After applying the cooling method the non--perturbative parts of the respective 
structure function have been fitted by an exponential function (with the same correlation 
length) over all distances between $0.1$ and $1\,\rm{fm}$. This contribution to 
${\cal D}$ is considerably bigger compared to ${\cal D}_1$.
This observation points towards a dominance of (anti)selfdual field entering the 
correlator (\ref{eq2}). In the one-instanton approximation ${\cal D}_1$ is exactly 
vanishing. The structure functions extracted from the cooled Monte Carlo
data still exhibit a perturbative tail $~\sim x^{-4}~$. In this case, however, the 
relative size of this contribution to ${\cal D}$ and ${\cal D}_1$ is not understood, 
in view of the above considerations.
  
In the lowest order (in $g^2$) semiclassical approximation for the field-strength 
correlator the Gaussian integral over quantum fluctuations above a single classical 
field configuration is separated from the gauge invariant product of field strengths 
to be evaluated for this background field. What then remains are zero--mode integrations 
over the appropriate set of collective coordinates (summarized as $~\Gamma~$ 
which characterizes the classical field) with a density function $~{\cal M}(\Gamma)$,
which results from the integration over non-zero mode fluctuations :
\begin{equation}
\label{Okt_25_1}
{\cal D}_{\mu\rho,\nu\sigma}(x)=
\frat{1}{Z}\int d\Gamma {\cal M}(\Gamma) \cdot
{\mathrm Tr}\{G_{\mu\rho}(x_1;\Gamma) S(x_1,x_2;\Gamma)
G_{\nu\sigma}(x_2;\Gamma) S^{\dagger}(x_1,x_2;\Gamma)\}\,,
\end{equation}
where $G_{\mu\nu}(x;\Gamma)$ is the field strength tensor and $S(x_1,x_2;\Gamma)$ 
the Schwinger line corresponding to configurations $A_\mu(x,\Gamma)$. 
To be more specific, we imagine a model of the vacuum state that is semi--classically 
represented by superpositions of $~N~$ instantons  and $~{\bar N}~$ anti-instantons
\cite{cal}

\begin{equation}
\label{Okt_25_2}
A_\mu(x,\Gamma)=\sum^{N}_{i=1}A_\mu(x;\gamma_i)+\sum^{\bar N}_{j=1}\bar
A_\mu(x;\bar \gamma_j)\,.
\end{equation}
The $\gamma_i$ ($~{\bar \gamma}_j~$) denote the collective
coordinates of the $i$-th instanton  ($j$-th anti-instanton), which include the
positions $z_i$, the group space orientations $\omega_i$, and the sizes $\rho_i$.
The integration measure in Eq. (\ref{Okt_25_1}) is then expressed by
$$
d\Gamma~=~\prod^{N}_{i=1} d \gamma_i
\prod^{\bar N}_{j=1} d \bar\gamma_j\,, \qquad
d \gamma_i~=~d^4 z_i d \omega_i d \rho_i\,, \qquad
d \bar\gamma_j~=~d^4 {\bar z}_j d \bar\omega_j d \bar\rho_j\,.
$$
For practical use we will consider here only instantons and antiinstantons of fixed 
size $\rho$. This corresponds to the instanton liquid model invented in \cite{shur}
with a delta-like size distribution. 
A more realistic $\rho$-distribution with a selfconsistent exponential
infra-red cutoff (allowing to satisfy low-energy theorems) can be
obtained from the assumption that (anti-)instantons repel each other at
short distances \cite{ilgmmp,diak}.\footnote{Recently, another interpretation 
has been put forward for such a shape as a manifestation of suppression by a Higgs 
like monopole condensate \cite{shuryak_size}.}

If the instanton liquid is sufficiently dilute we can approximate
the functional integral by an expansion in powers of the
(anti-)instanton densities $~n_4~=N/V$ ($~{\bar n_4={\bar N}/V}~$). Then it is natural 
first to try to neglect possible correlation effects due to interactions between instantons.

Strictly speaking, the superposition ansatz (\ref{Okt_25_2}) makes sense as an approximate 
saddle point of the action only if the vector potentials $A_\mu,$ ${\bar A}_\mu$ decrease 
fast enough. This happens when the singular gauge expression is used for the
(anti-) instanton solutions $A_\mu,$ ${\bar A}_\mu$, instead of
the regular gauge form \cite{cal}.  The existence of a systematic expansion in higher 
order contributions to the measure ${\cal M}(\Gamma)$ has been proven in Ref. \cite{levine}.

In \cite{IMMMPS} the leading term in an expansion of (\ref{eq1}) in terms of the 
density has been discussed in detail. In that approximation the field strength correlator 
is given by the sum of instanton ($I$) and antiinstanton (${\bar I}$) contributions
\begin{eqnarray}
\label{eq8} 
{\cal D}^{(1)}_{\mu\rho,\nu\sigma}(x_1,x_2)
& = & {\cal D}^{I}_{\mu\rho,\nu\sigma} + {\cal D}^{\bar I}_{\mu\rho,\nu\sigma}
\nonumber \\
& = & n_4 \int d^4 z~
{\mathrm Tr} \left\{ G_{\mu\rho}(x_1;\gamma) S(x_1,x_2;\gamma) G_{\nu\sigma}(x_2;\gamma)
S^{\dagger}(x_1,x_2;\gamma) \right\} \\
&& + (n_4,\gamma \to \bar n_4,\bar\gamma)\,.
\nonumber
\end{eqnarray}
The integration over the global group orientation of the respective solution is trivial 
in this case and can be omitted. The Schwinger line phase factor is a path dependent 
matrix in the fundamental representation,
\begin{equation}
\label{eq9}
S(x_1,x_2;z)=P\;\exp\left( i \int^{1}_{0}d t\, \dot{x}_\mu(t)
A_\mu(x(t);z)\right)\,,
\end{equation}
where the vector potential $A_\mu=  T^a A^a_{\mu}$, in the case at hand belongs 
to the single instanton source localized at $z$.
Advantage has been taken from the fact that this phase factor, evaluated for a straight 
line has a closed expression which can be easily contracted with the field strengths.

As mentioned before, due to the vanishing of ${\cal D}_1$ for purely (anti)selfdual fields,
the single-instanton approximation reads equally for the longitudinal and transverse 
correlators
\begin{equation}
\label{firstorder}
{\cal D}_{||}(x^2) = {\cal D}_{\bot}(x^2) = \frac{4}{3} \pi^2 n I(\frac{x}{\rho})\,,
\end{equation}
where $n=n_4 + \bar n_4$ is the total density of instantons plus antiinstantons.
The function $I(x/\rho)~$ (for a {\it straight line} Schwinger phase
factor) is normalized by $I(0)=1$. It has been numerically computed in \cite{IMMMPS} 
in order to deal with the final integration over the instanton position $z$ and is 
plotted in Fig. 1.

\section{Revisiting the Second Order Instanton Density  \\ Contributions 
to the Field Strength Correlators}
%===================================================================

In this section we shall present  the numerical integration giving the next order
term in an expansion in terms of the instanton density. Now we have to consider 
the field strength from the nonlinear superposition of two different solutions $A$ and $B$,
where both $A$ and $B$ can represent an instanton or antiinstanton,
\begin{eqnarray}
\label{Okt_25_3}
G_{\mu\rho}(A,B)&=&G_{\mu\rho}(A)+G_{\mu\rho}(B)+
\Delta G_{\mu\rho}(A,B)\,,\nonumber\\ [-.2cm]
\\ [-.25cm]
\Delta G_{\mu\rho}(A,B)&=&-i  \{[A_\mu, B_\rho] +
[B_\mu,A_\rho] \}\,.\nonumber
\end{eqnarray}

\noindent
This field strength is plugged into an integral over the (factorized) 
two--source weight function (with fixed sizes $\rho_1=\rho_2=\rho$)
\begin{eqnarray}
\label{SecOrd}
{\cal D}_{\mu\rho,\nu\sigma}^{(2)}(x_1,x_2)  =
\frac{1}{2} \sum_{A,B=I,{\bar I}}~n_4^{(A)}~n_4^{(B)}
\int d^4 z_1 \int d^4 z_2~
\int d \omega_1 \int d \omega_2~
\times \nonumber \\ [-.2cm]
 \\ [-.25cm]
{\mathrm Tr} \left\{
G_{\mu\rho}(A(x_1,\gamma_1),B(x_1,\gamma_2))S(x_1,x_2;\gamma_1,\gamma_2)~
G_{\nu\sigma}(A(x_2,\gamma_1),B(x_2,\gamma_2))S^{\dagger}(x_1,x_2;\gamma_1,\gamma_2) 
   \right\}\,. \nonumber
\end{eqnarray}

\noindent
For the purpose of our calculation, the points $x_1$ and $x_2$ are sitting on the 
Euclidean time axis while the two (anti)instantons  are arbitrarily located in 
4-d space-time. One has to keep in mind, that those formal contributions to Eq. 
(\ref{SecOrd}), where both the factors $G_{\mu\nu}(A,B)$ at $x_1$ and $x_2$ would 
receive contributions from only one and the same $G_{\mu\nu}(A)$ (or $G_{\mu\nu}(B)$) 
do not really occur. They are already taken into account in the first order contribution 
${\cal D}^{(1)}$ (\ref{eq8}).

\medskip\noindent
With the notation  $y_1=x-z_1$, $y_2=x-z_2$ the two instanton vector potentials, 
both in the singular gauge, read as follows
\begin{eqnarray}
\label{SecOrd_sng}
A_\mu^{a}(x;z_1)~ =~  \bar\eta_{a \mu \nu}~
y_{1\nu} f(y_1,\rho)\,,~B_\mu^{a}(x;z_2)~ =~ \omega_{aa'}\bar\eta_{a'\mu\nu}~
 y_{2\nu} f(y_2,\rho)\,,~~f(y,\rho)= \frat{2~\rho^2}{y^2(y^2+\rho^2)}\,,
\end{eqnarray}
where $\omega_{aa'}$ is included to take the relative color orientation inside the pair 
into account. $\bar \eta_{a\mu\nu}$ and $\eta_{a\mu\nu}$ are the 't Hooft tensors 
\cite{thooft}. For an antiinstanton instead of an instanton replace $\bar \eta \to \eta$.

The rotation of instantons in color space (the case of $SU(2)$ group)
$$
A^a_\mu \rightarrow \omega^{ab} A^b_\mu
$$
can be represented by the related rotation in $3-d$ coordinate space
$$
\omega^{ab} A^b_i(x_0,{\vf x})= \omega^{-1}_{ij} A^a_j(x_0,\omega{\vf x})\,,
$$
$$
\omega^{ab} A^b_4(x_0,{\vf x})=A^a_4(x_0,\omega{\vf x})\,.
$$
This property will be very important in the further average over the relative
color orientation of two instantons. 

First, let us consider the correlator ${\cal D}_{||}(x^2)$. According to Eqs. 
(\ref{Okt_25_3},\ref{SecOrd}), the evaluation of (\ref{dpar1}) 
naively gives rise to 16 terms. After averaging over the relative color and $3-d$ space
orientations of two instantons only 
six
of these terms give a nonzero
contribution. Let us describe the contributions in some detail.
\begin{enumerate}
\item
Two terms, for which both the $G_{4i}(A(x_1,\gamma_1),B(x_1,\gamma_2))$ 
and the $G_{4i}(A(x_2,\gamma_1),B(x_2,\gamma_2))$ 
are represented by one and the same $G_{4i}(A)$ (or $G_{4i}(B)$) do not really
occur (see above). 
\item
If in the place of $G_{4i}(A(x_1,\gamma_1),B(x_1,\gamma_2))$ the term 
$G_{4i}(A)$ is taken (or $G_{4i}(B)$) and in the place of 
$G_{4i}(A(x_2,\gamma_1),B(x_2,\gamma_2))$ the term $G_{4i}(B)$ (or $G_{4i}(A)$, 
respectively) the resulting contributions (2 terms) are identically zero. 
This can be easily seen. Under color rotation of the second instanton $B$
(or $A$, respectively) (being equivalent to a $3-d$ rotation) the term in place of
$G_{4i}(A(x_2,\gamma_1),B(x_2,\gamma_2))$ becomes
rotated, while that in place of $G_{4i}(A(x_1,\gamma_1),B(x_1,\gamma_2))$ does not.
The rotation of the coordinates of the second instanton in the phase factor
$S(x_1,x_2;\gamma_1\gamma_2)$ etc. can be skipped, because there is an integration 
over these coordinates. As a result we have to average over the $3-d$ vector 
$G_{4i}(B)$ (or $G_{4i}(A)$) orientation leading to a vanishing contribution.
\item
If in the place of $G_{4i}(A(x_1,\gamma_1),B(x_1,\gamma_2))$ the term $G_{4i}(A)$
is inserted and in the place of $G_{4i}(A(x_2,\gamma_1),B(x_2,\gamma_2))$ the term 
$-i  [A_4, B_i]$, the contribution is zero by the same reason as in the 
previous consideration. There are three other vanishing contributions to be
obtained from the described one by the replacement
$A \leftrightarrow B$, $x_1 \leftrightarrow x_2$ (4 terms).
\item
If in the place of $G_{4i}(A(x_1,\gamma_1),B(x_1,\gamma_2))$ the term $G_{4i}(A)$
is inserted and in the place of $G_{4i}(A(x_2,\gamma_1),B(x_2,\gamma_2))$ the term 
$-i  [B_4, A_i]$ the corresponding contribution is nonzero. It is independent of
the color orientation of the second instanton $B$ (equivalent to
its rotation in $3-d$ space) as far as the rotation of the second instanton
in the phase factor $S(x_1,x_2;\gamma_1\gamma_2)$ etc. is integrated out.
There are three other analogous terms with nonzero contribution which can be
obtained from the described one by the replacements
$A \leftrightarrow B$, $x_1 \leftrightarrow x_2$ (4 terms).
\item
If in the place of $G_{4i}(A(x_1,\gamma_1),B(x_1,\gamma_2))$ the term $-i  [A_4, B_i]$ 
(or $-i [B_4,A_i]$) is inserted and in the place of $G_{4i}(A(x_2,\gamma_1),B(x_2,\gamma_2))$ 
the term $-i  [B_4,A_i]$ (or $-i  [A_4,B_i]$, respectively),  the contribution is vanishing
by the same reason as discussed in points 2, 3 above (2 terms).
\item
Finally, if in the place of $G_{4i}(A(x_1,\gamma_1),B(x_1,\gamma_2))$ the term $-i  [A_4,B_i]$
(or $-i  [B_4,A_i]$) is inserted and in the place of $G_{4i}(A(x_2,\gamma_1),B(x_2,\gamma_2))$  
the term  $-i [A_4,B_i]$  (or $-i  [B_4,A_i]$, respectively), the contribution is nonvanishing 
and independent  of color rotation of the second instanton $B$ (equivalent to its 
rotation in $3-d$ space) by the same reason as in point 4 above (2 terms).
\end{enumerate}
Now let us consider the correlator ${\cal D}_{\bot}(x^2)$.
The corresponding expression in (\ref{dpar1}) naively 
contains 16 terms upon insertion of (\ref{Okt_25_3}) into Eq. (\ref{SecOrd}).
After averaging over the relative color and $3-d$ space
orientations of two instantons only 
four
of them give nonzero contribution.
\begin{enumerate}
\item
Two terms, for which both the $G_{ij}(A(x_1,\gamma_1),B(x_1,\gamma_2))$ 
and the $G_{ij}(A(x_2,\gamma_1),B(x_2,\gamma_2))$ 
are represented by one and the same $G_{ij}(A)$ (or $G_{ij}(B)$) do not really
occur (see above). 
\item
If in the place of $G_{ij}(A(x_1,\gamma_1),B(x_1,\gamma_2))$ the term $G_{ij}(A)$ is 
taken  (or $G_{ij}(B)$) and in the place of $G_{ij}(A(x_2,\gamma_1),B(x_2,\gamma_2))$ the 
term $G_{ij}(B)$ (or $G_{ij}(A)$, respectively) the contribution is zero. 
Under color rotation of the second instanton $B$ (or $A$, respectively)
which is equivalent to its rotation in $3-d$ space
$G_{ij}(A(x_2,\gamma_1),B(x_2,\gamma_2))$ is rotated,
while $G_{ij}(A(x_1,\gamma_1),B(x_1,\gamma_2))$ is not. The rotation of
the coordinates of the second instanton in the phase factor
$S(x_1,x_2;\gamma_1\gamma_2)$ etc. can be skipped, because of the integration
over these coordinates. As a result we have the average over the
orientation of a $3-d$ vector (an antisymmetric tensor is equivalent
to a $3-d$ vector) $G_{ij}(B)$ (or $G_{ij}(A)$), which gives zero
(2 terms).
\item
If in the place of $G_{ij}(A(x_1,\gamma_1),B(x_1,\gamma_2))$ the term $G_{ij}(A)$ is 
inserted and in the place of $G_{ij}(A(x_2,\gamma_1),B(x_2,\gamma_2))$ the term 
$-i  [A_i, B_j]$ the contribution is zero by the same reason as in point 2. There 
are three other terms with zero contribution which can be obtained from the described
one by the replacements $A \leftrightarrow B$, $x_1\leftrightarrow x_2$
(4 terms).
\item
If in the place of $G_{ij}(A(x_1,\gamma_1),B(x_1,\gamma_2))$ the term $G_{ij}(A)$ is 
inserted and in the place of $G_{ij}(A(x_2,\gamma_1),B(x_2,\gamma_2))$ the term 
$-i  [B_i, A_j]$ the contribution is zero by the same reason as in point 2. There 
are three other terms with zero  contribution which can be obtained from the described
one by the replacements $A \leftrightarrow B$, $x_1\leftrightarrow x_2$
(4 terms).
\item
If in the place of $G_{ij}(A(x_1,\gamma_1),B(x_1,\gamma_2))$ the term $-i  \{[A_i, B_j]+
[B_i, A_j]\}$ (at $x_1$) is inserted and in the place of 
$G_{ij}(A(x_2,\gamma_1),B(x_2,\gamma_2))$ the same term 
$-i  \{[A_i, B_j]+ [B_i, A_j]\}$ (at $x_2$) the contribution is non-vanishing (4 terms).
\end{enumerate}

With the second order contributions in terms of the instanton density included
and with an additional (unknown) perturbative short-range $x^{-4}$-contribution the 
longitudinal and transverse correlators can be directly expressed  as follows

\begin{eqnarray}
\label{dpar2}
{\cal D}_{||}(x^2) = \frat{4}{3} \pi^2 n I(\frac{x}{\rho}) + 
		     \frat{27}{16} \pi^4 n^2 \rho^4 I_{||}(\frac{x}{\rho}) + 
		     \frat{a_{||}}{x^4}\,,
\nonumber\\ ~~~\\
{\cal D}_{\bot}(x^2) = \frat{4}{3} \pi^2 n I(\frac{x}{\rho}) + 
		       \frat{27}{16} \pi^4 n^2 \rho^4 I_{\bot}(\frac{x}{\rho}) + 
		       \frat{a_{\bot}}{x^4}\,,\nonumber
\end{eqnarray}
where $n$ is the total density of instantons and antiinstantons and
$\rho$ is their size, $a_{||}$ and $a_{\bot}$ represent the coefficients of the respective
perturbative contributions. 

The functions $I(x/\rho)$, $I_{||}(x/\rho)$ and $I_{\bot}(x/\rho)$ have been obtained by
numerical integration and are all normalized to 1 at $x = 0$. The numerical 
factors come from the transition to the
$SU(3)$ gauge group into which the $SU(2)$-instantons are embedded and can be calculated 
analytically by the consideration of correlators at $x = 0$.
There is a twofold nontrivial integration in $I(x/\rho)$, and a fivefold integration 
is implied by the expressions for $I_{||}(x/\rho)$ and $I_{\bot}(x/\rho)$. In the last
case the integrations have been carried out by means of a Monte Carlo
importance sampling method. The two points $x_1$ and $x_2$ of the correlator, 
with $x=|x_1-x_2|$, are located on the Euclidean time axis, at times $\pm x/2$.
The two-instanton (or instanton-antiinstanton) contribution is
obtained by integrating over $r_1$, $t_1$, $r_2$, $t_2$ and $\theta$ 
in a sequential manner, where the positions of the two instanton
centers are $x^{I}_{1}=(0, 0, r_1, t_1)$ and 
$x^{I}_{2}=(r_2~\sin\theta, 0, r_2~\cos\theta, t_2)$.
First, variables $(r_1, t_1)$ were generated in a box of size
$(20~\rho, 40~\rho)$ with a distribution proportional to
$r_1^2~f(x_{11})~f(x_{12})$, where $f(x)=\frac{2}{x}~\frac{\rho^2}{(x^2+\rho^2)}$ 
is the profile function of an instanton. Here, $x_{11}$ and $x_{12}$ are the $4-d$ distances 
of the first instanton $x^{I}_{1}$ from $x_1$ and $x_2$, respectively.
For this sampling of $(r_1, t_1)$ the acceptance varied from $1/1000$ to $1/100$ for 
$0 < x < 5~\rho$. About $7000$ events were accepted.
Second, for each accepted event new variables $(r_2, t_2)$ were
generated in a similar way, and about $100$ events were accepted per 
each $x^{I}_{1}=(0, 0, r_1, t_1)$. 
Finally, for each of the events accepted so far, an angle $\theta$ of relative orientation in
$3-d$ space has been randomly selected 100 times, according to a flat measure in $cos~\theta$. 
The obtained accuracy of the Monte Carlo integration was about 1~\%.   
The convergence of integration has been verified by doubling the integration
box for $(r_1, t_1)$ and $(r_2, t_2)$ to $(40~\rho, 80~\rho)$.

The functions $I_{||}(x/\rho)$ and $I_{\bot}(x/\rho)$
are plotted together with $I(x/\rho)$ obtained in Ref. \cite{IMMMPS}
in Fig. 1. The longitudinal and transverse  correlators from
Eqs. (\ref{dpar2}) have been jointly fitted to the lattice data of
Ref. \cite{dig2} within a distance range from $0.4\,\rm{fm}$ to $1\,\rm{fm}$.
For the parameters $n$, $\rho$, $a_{||}$ and $a_{\bot}$ we have found
$$ n = (1 \pm 0.1)\,\rm{fm}^{-4}\,, \quad \rho = (0.42 \pm 0.01)\,\rm{fm}\,,
\quad a_{||} = 0.46 \pm 0.02\,, \quad a_{\bot} = 0.76 \pm 0.06 $$
with an $\chi^2 / N_{\rm{dof}} = 56.1 / (30-4)$.
The corresponding curves for the correlations functions ${\cal D}_{||}$ and ${\cal D}_{\bot}$
are drawn in Figs. 2 and 3, respectively, together with the lattice data.
In order to give an impression how large the contribution of the
different terms in Eq. (\ref{dpar2}) are, we have plotted them separately.

\section{Conclusions and Discussion}
%===================================
We have considered the two-point correlators of gluon field strengths in the
uncorrelated instanton liquid model up to the second order in the instanton density.
The correlators have two parts: a nonperturbative one which is considered exclusively 
due to instantons and a perturbative one due to gluon exchange 
fluctuations in the vacuum. We have fitted the resulting expression
directly to the lattice data within a restricted distance range.
For this range the achieved quality of our fit is comparable with those
of \cite{dig2}, where purely exponential terms together with
$x^{-4}$-contributions were fitted to the lattice data after a separation into 
${\cal D}$ and ${\cal D}_1$. The value for the instanton density comes near to the value
expected from phenomenological applications, although our present analysis describes 
quenched lattice data. There is a recent analysis due to A. Hasenfratz 
\cite{a_hasenfratz} which, on the basis of the  two-point correlator
of the topological density, gives estimates for the instanton density, 
the fraction of two-instanton and instanton-antiinstanton molecules, all for 
quenched $SU(3)$ theory and full QCD. The instanton density was there found to be about
$1\,\rm{fm}^{-4}$ for pure $SU(3)$, too, however the instanton size was estimated as 
about $0.3\,\rm{fm}$. Thus, the mean instanton size obtained here is more in accordance 
with UKQCD \cite{UKQCD}. The packing fraction of the instanton gas is estimated to
$n\rho^4 \simeq .03$ which corresponds to a reasonable diluteness.

In our fit for the range $0.4\,\rm{fm} < x < 1\,\rm{fm}$  the instanton and
perturbative contributions are of a comparable order of magnitude.
We obtain the relation ${\cal D}_{\bot} > {\cal D}_{||}$ now both for the perturbative and 
the instanton contributions. This means that both contributions to the invariant function
${\cal D}_1$ turn out to be positive. This is contrary to our previous result in \cite{IMMMPS}
where we concluded that ${\cal D}_1 < 0$ is inavoidable within the instanton liquid description.
The neglect of the Schwinger-line phase factor in the two-instanton (instanton-antiinstanton 
field) is to be blamed for this wrong conclusion.

The second order density terms are the final ones for transverse correlators, 
{\it i.e.} the  $O(n^3), O(n^4), \cdots$ terms would be vanishing.
For the longitudinal correlators there exist also a non-vanishing $O(n^3)$ contribution, 
but no $O(n^4)$ ones. The calculation of the longitudinal correlators up to this order 
require a nine-dimensional nontrivial integration over the three-instanton degrees of freedom.
We hope to come back to this integration in future.

\section*{Acknowledgements}
%=========================
The authors are grateful to S.~V.~Molodtsov and
Yu.~A.~Simonov for useful discussions.
The financial support through the joint RFFI-DFG project
436 RUS 113/309/10 (R) is gratefully acknowledged.

\newpage
%======================================================
%  Figures
%======================================================
\begin{figure}
\begin{center}
\leavevmode
\epsfxsize = 12cm
\epsffile[40 50 540 690]{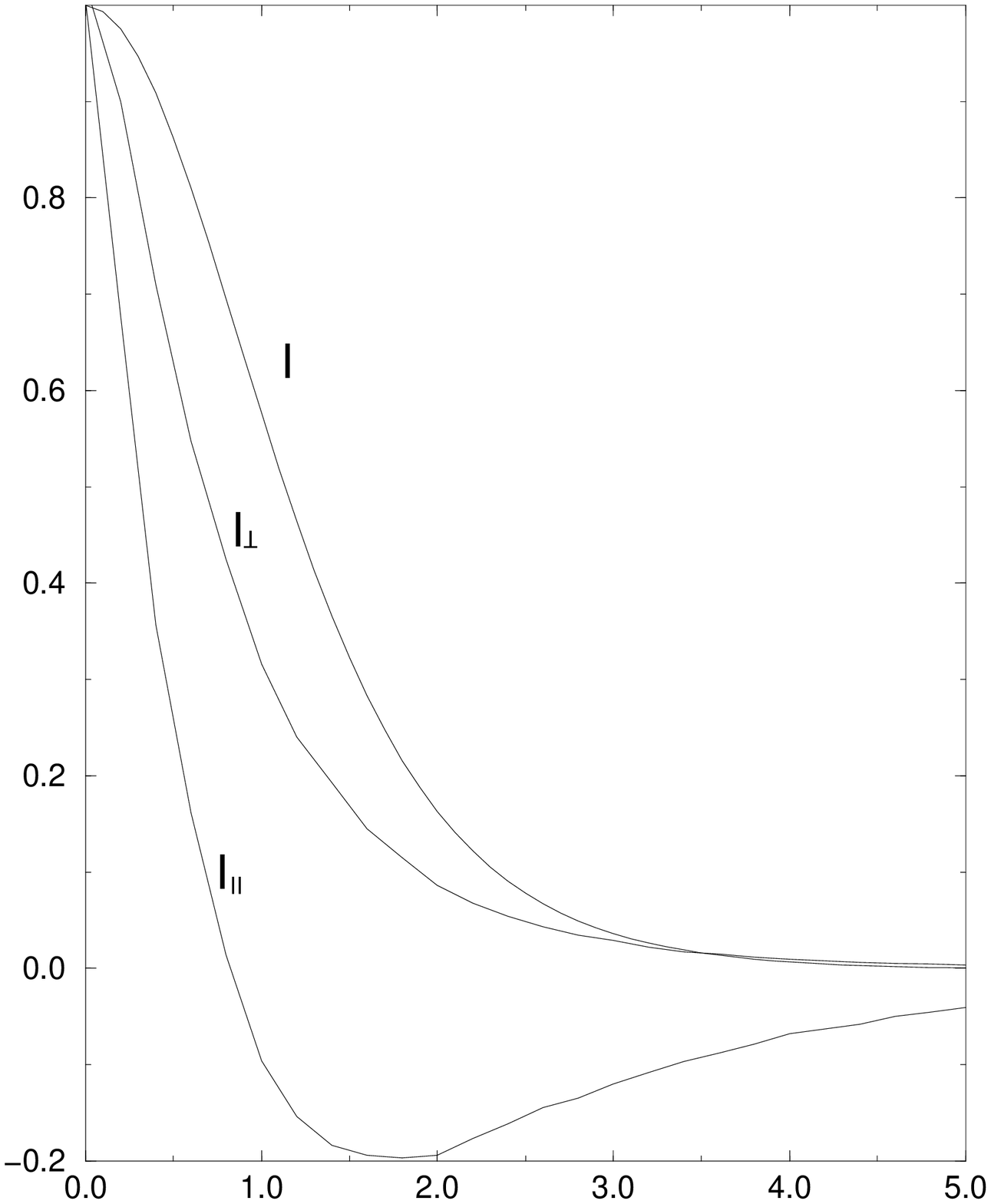}
\label{fig1}
\caption{The numerically computed functions 
$I(x/\rho)$ for the first order instanton contribution 
(from [31]) and $I_{||}(x/\rho)$, $I_{\bot}(x/\rho)$ 
for the second order contribution.}
\end{center}
\end{figure}
%=======================================================
\begin{figure}
\begin{center}
\leavevmode
\epsfxsize = 12cm
\epsffile[40 50 540 690]{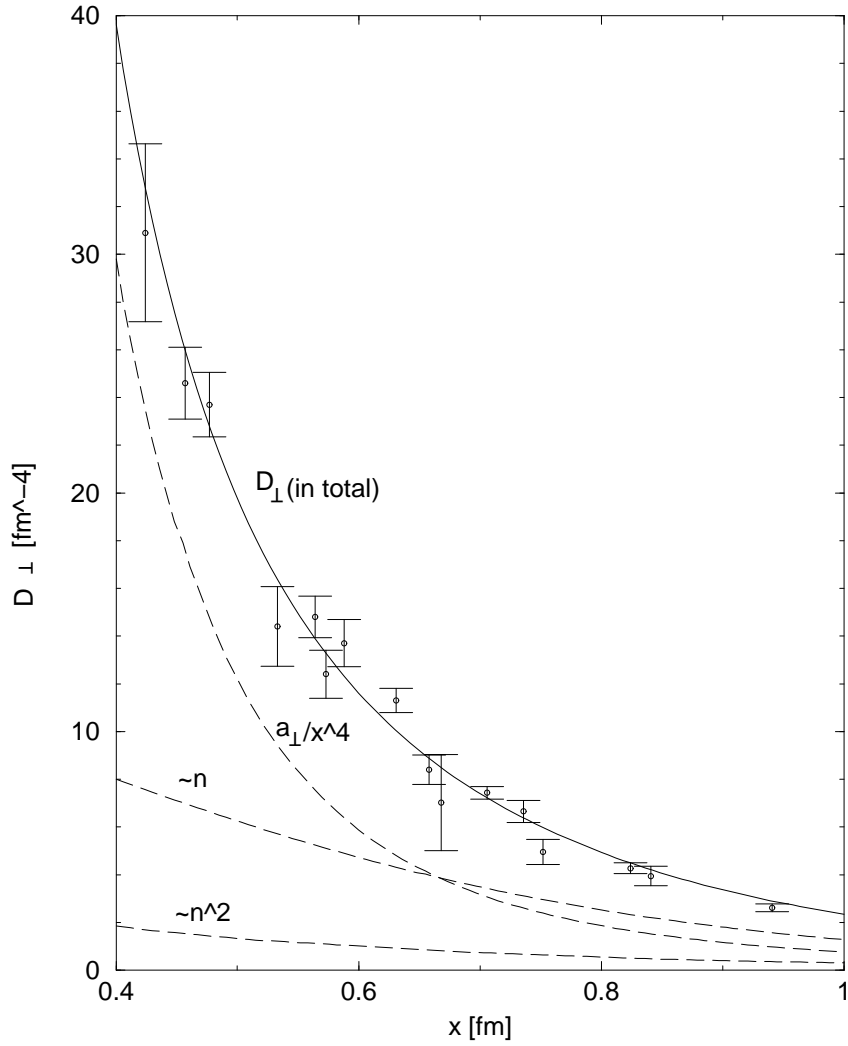}
\label{fig2a}
\caption{Transverse field strength correlator as described
by the instanton gas model (continuous line) and fitted to the
lattice data points of Ref. [9] The dashed lines show
separately the first and second order instanton density contributions
as well as the perturbative $x^{-4}$ contribution.}
\end{center}
\end{figure}
%=======================================================
\begin{figure}
\begin{center}
\leavevmode
\epsfxsize = 12cm
\epsffile[40 50 540 690]{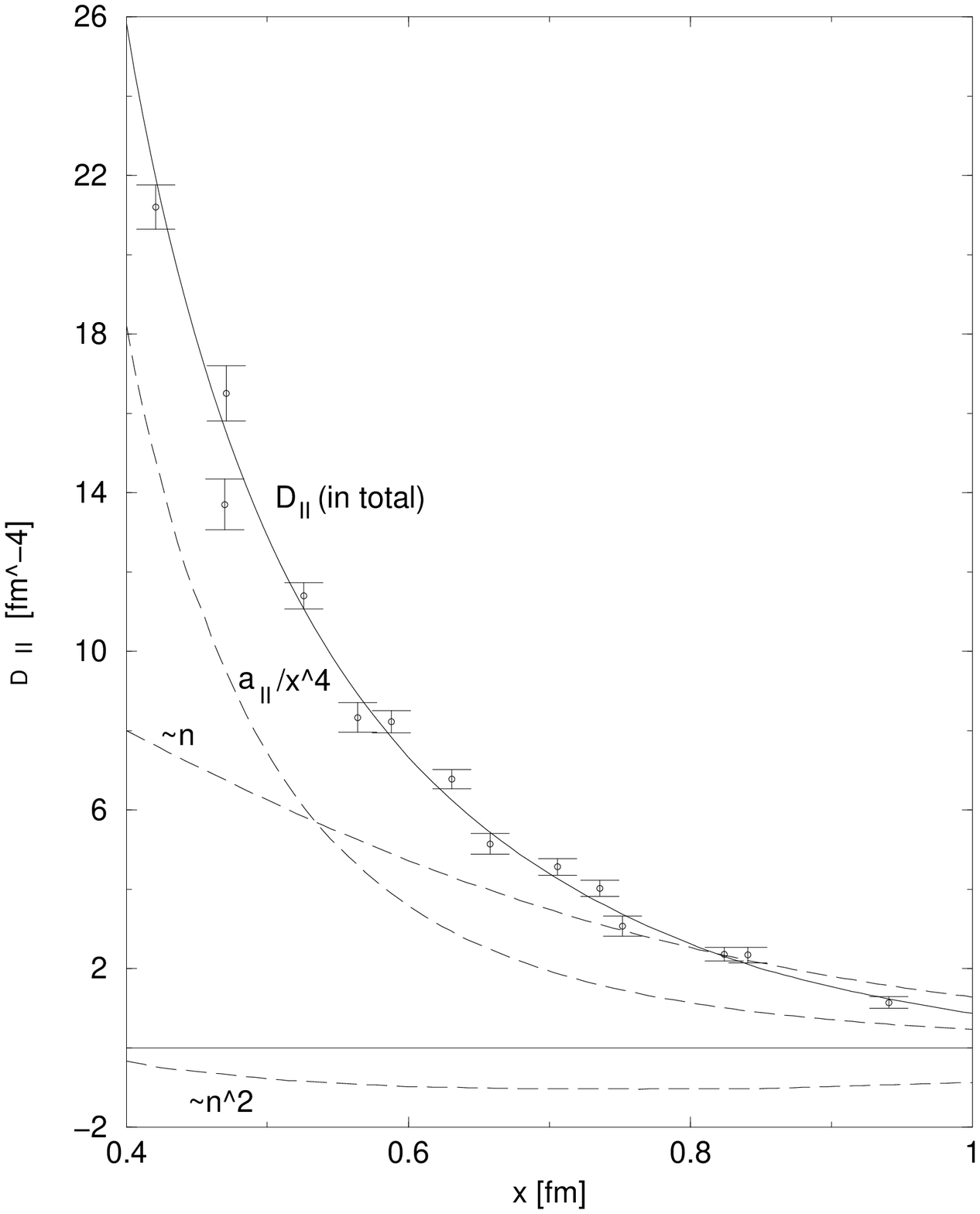}
\label{fig2b}
\caption{Same as in Figure \ref{fig2a}, but for the longitudinal
field strength correlator.}
\end{center}
\end{figure}
%=======================================================

\begin{thebibliography}{99}
%==========================
%
\bibitem{simdosch}
H. G. Dosch, Phys. Lett. {\bf B190} (1987)  177; \\
H. G. Dosch and Yu. A. Simonov, Phys. Lett. {\bf B205} (1988)  339; \\
Yu. A. Simonov, Nucl. Phys. {\bf B324} (1989)  67.
%
\bibitem{sim}
Yu. A. Simonov, Usp. Fiz. Nauk {\bf 166} (1996) 337.
%
\bibitem{simshev}
V. I. Shevchenko and Yu. A. Simonov,
Yad Fiz. {\bf 60} (1997) 1329; transl. Phys. At. Nucl. {\bf 60} (1997) 1201.
%
\bibitem{quarkonia}
Yu. A. Simonov, S. Titard, and F. J. Yndurain, Phys. Lett. {\bf B354} (1995) 435.
\bibitem{nachtmann}
H. G. Dosch, E. Ferreira, and A. Kr\"amer, Phys. Rev. {\bf D50} (1994)  1992; \\
O. Nachtmann, Schladming lectures 1996; \\
E. R. Berger and O. Nachtmann, Eur. Phys. J. {\bf C7} (1999) 459. 
%
\bibitem{campos}
M. Campostrini, A. Di Giacomo, and G. Mussardo, Z. Phys.  {\bf C25} (1984)  173.
%
\bibitem{dig1}
A. Di Giacomo and H. Panagopoulos, Phys. Lett. {\bf B285} (1992)  133.
%
\bibitem{ddeb}
L. Del Debbio, A. Di Giacomo, and Yu. A. Simonov, Phys. Lett. {\bf B332} (1994) 111.
%
\bibitem{dig2}
A. Di Giacomo, E. Meggiolaro, and H. Panagopoulos, e-Print Archive: hep-lat/9603017, and 
Nucl. Phys. Proc. Suppl. {\bf 54A} (1997) 343.
%
\bibitem{dig3}
A. Di Giacomo, E. Meggiolaro, and H. Panagopoulos, Nucl. Phys. {\bf B483} (1997) 371.
%
\bibitem{dig4}
M. D'Elia, A. Di Giacomo, and E. Meggiolaro, Phys. Lett. {\bf B408} (1997) 315.
%
\bibitem{bali}
G. Bali, N. Brambilla, and A. Vairo, Phys. Lett. {\bf B421} (1998) 265-272.
%
\bibitem{DG_talk}
A. Di Giacomo, Lectures given at 
AUTUMN 99, Lisbon, Portugal, 29 Sep - 4 Oct 1999, e-Print Archive: hep-lat/9912016.
%
\bibitem{antonov}
D. Antonov, e-Print Archive: hep-ph/0001193.
% 
\bibitem{meggiolaro}
E. Meggiolaro, Phys. Lett. {\bf B451} (1999) 414. 
%
\bibitem{thooft}
G.'t Hooft, Phys. Rev. {\bf D14} (1976) 3432.
%
\bibitem{cal}
C. G. Callan, R. Dashen, and D. J. Gross,
Phys. Rev. {\bf D17} (1978) 2717. 
%
\bibitem{ilgmmp}
E.-M. Ilgenfritz and M. M\"uller-Preussker, Nucl. Phys. {\bf B184} (1981) 443.
%
\bibitem{shur}
E. V. Shuryak, Nucl. Phys.  {\bf B203} (1982) 93, 116, 140; 
Nucl. Phys.  {\bf B328} (1989) 85, 102.
%
\bibitem{diak}
D. I. Diakonov and V. Yu. Petrov, Nucl. Phys. {\bf B245} (1984) 259;\\
D. I. Diakonov, V. Yu. Petrov, and P. V. Pobylitsa, Phys. Lett.  {\bf B226} (1989) 471.
\bibitem{diaktalk}
D. I. Diakonov, Talk given at 
Varenna 1995,  e-Print Archive: hep-ph/9602375.
%
\bibitem{shurschaf}
T. Sch\"afer and E. V. Shuryak, Rev. Mod. Phys. {\bf 70} (1998) 323-426.
%
\bibitem{negele}
J. W. Negele, Nucl. Phys. Proc. Suppl. {\bf 73} (1999) 92.
% 
\bibitem{teper}
M. Teper, Plenary talk 
LATTICE 99, Pisa, Italy, 29 Jun - 3 Jul 1999. e-Print Archive: hep-lat/9909124. 
\bibitem{fukushima}
M. Fukushima, H Suganuma, and H. Toki, Phys. Rev. {\bf D60} (1999) 094504.
\bibitem{chen}
D. Chen, R. C. Brower, J. W. Negele, and E. V. Shuryak,
Nucl. Phys. Proc. Suppl. {\bf 73} (1999) 512.
%
\bibitem{ilgthur}
E.-M. Ilgenfritz and S. Thurner, e-Print Archive: hep-lat/9810010.
%
\bibitem{monopolperc}
M. Fukushima, S. Sasaki, H. Suganuma, A. Tanaka, H. Toki, and D. Diakonov,
Phys. Lett. {\bf B399} (1997) 141.
%
\bibitem{ba}
V. N. Baier and Yu. F. Pinelis, Phys. Lett. {\bf B116} (1982)  179.
%
\bibitem{do1}
A. E. Dorokhov, S. V. Esaibegyan, and S. V. Mikhailov,
Phys. Rev. {\bf D56} (1997) 4062.
%
\bibitem{IMMMPS}
E.-M. Ilgenfritz , B. V. Martemyanov , S. V. Molodtsov, M. M\"uller--Preussker,
and Yu. A. Simonov , Phys. Rev. {\bf D58} (1998) 114508.
%
\bibitem{do2}
A. E. Dorokhov, S. V. Esaibegyan, A. E. Maximov, and S. V. Mikhailov,
e-Print Archive: hep-ph/9903450.
%
\bibitem{sim2}
V. I. Shevchenko and Yu. A. Simonov, Phys. Lett. {\bf B437} (1998) 146-152.
%
\bibitem{shuryak_size}
E. V. Shuryak, e-Print Archive: hep-ph/9909458.
%
\bibitem{levine}
H. Levine and L. G. Yaffe, Phys. Rev. {\bf D19} (1979) 1225.
%
\bibitem{a_hasenfratz} A. Hasenfratz, 
e-Print Archive: hep-lat/9912053.
%
\bibitem{UKQCD}
D. A. Smith and M. J. Teper, Phys. Rev. {\bf D58} (1998) 014505.
\end{thebibliography}
 \end{document}